\documentclass[a4paper,preprintnumbers,floatfix,superscriptaddress,pra,twocolumn,showpacs]{revtex4}

\usepackage{amsmath, amsthm, amssymb}
\usepackage{graphicx}
\usepackage{dcolumn}
\usepackage{bm}
\usepackage{hyperref} 
\usepackage{braket}
\usepackage{color}
\usepackage{dsfont}
\usepackage{multirow}
\usepackage{booktabs}
\newcommand{\ra}[1]{\renewcommand{\arraystretch}{#1}}
\newcommand{\figref}[1]{Fig.~\ref{#1}}
\newcommand{\tabref}[1]{Table~\ref{#1}}
\newcommand{\secref}[1]{Sec.~\ref{#1}}
\newcommand{\appref}[1]{App.~\ref{#1}}
\begin{document}

\title{Experimental multipartite entanglement and randomness certification
of the W state in the quantum steering scenario}

\author{A. M\'{a}ttar}\altaffiliation[Corresponding author.\,\,\,]{alejandro.mattar@icfo.es}\affiliation{ICFO-Institut de Ciencies Fotoniques, The Barcelona Institute of Science and Technology, 08860 Castelldefels (Barcelona), Spain}
\author{P. Skrzypczyk}\affiliation{H. H. Wills Physics Laboratory, University of Bristol, Tyndall Avenue, Bristol, BS8 1TL, United Kingdom}
\author{G. H. Aguilar}\affiliation{Instituto de F\'{i}sica, Universidade Federal do Rio de Janeiro, CP 68528, 21941-972, Rio de Janeiro, RJ, Brazil}\affiliation{Institute for Quantum Science and Technology, and Department of Physics and Astronomy, University of Calgary, Calgary, Alberta T2N 1N4, Canada}
\author{ R. V. Nery}\affiliation{Instituto de F\'{i}sica, Universidade Federal do Rio de Janeiro, CP 68528, 21941-972, Rio de Janeiro, RJ, Brazil}
\author{ P. H. Souto Ribeiro}\affiliation{Departamento de F\' \i sica, Universidade Federal de Santa Catarina, CEP 88040-900, Florian\'opolis, SC,  Brazil}
\author{ S. P. Walborn}\affiliation{Instituto de F\'{i}sica, Universidade Federal do Rio de Janeiro, CP 68528, 21941-972, Rio de Janeiro, RJ, Brazil}
\author{D. Cavalcanti}\affiliation{ICFO-Institut de Ciencies Fotoniques, The Barcelona Institute of Science and Technology, 08860 Castelldefels (Barcelona), Spain}

\date{\today}

\begin{abstract}
Recently [Cavalcanti \textit{et al.} Nat Commun \textbf{6}, 7941 (2015)] proposed a method to certify the presence of entanglement in asymmetric networks, where some users do not have control over the measurements they are performing. Such asymmetry naturally emerges in realistic situtations, such as in cryptographic protocols over quantum networks. Here we implement such ``semi-device independent'' techniques to experimentally witness all types of entanglement on a three-qubit photonic W state. Furthermore we analise the amount of genuine randomness that can be certified in this scenario from any bipartition of the three-qubit W state.
\end{abstract}

\maketitle

\section{Introduction} \label{sec.intro}

In general, the certification of \textit{entanglement} \cite{Schrodinger35,Horodecki2009} requires to perform specific measurements on the quantum state one is willing to test \cite{Guhne2009}. Experimentally, this is a delicate task since mismatches between either the state or the measurements and their actual physical implementation may lead to false-positive conclusions about the presence of entanglement in the network \cite{Rosset2012}. Although such mismatches can in principle be patched in experimental setups, the situation becomes dramatic when considering applications, where the devices used are not trusted as they could have been provided and controlled by some adversary. One solution to this problem is the use of \textit{device-independent} techniques \cite{BrunnerCavalcantiPironioScaraniWehner2013,acin2007}, for which no assumption is made on the devices that generate the state or perform the measurements. In this approach the devices are seen as black boxes, accessed only with classical inputs (corresponding to the measurement choices) and providing classical outputs (corresponding to the measurement outcomes). Although such \textit{device-independent entanglement witnesses} have been soundly considered in the past years \cite{Bancal2011, Barreiro2013}, their physical implementation turns out to be very demanding \cite{Gerhardt2011} for it requires one to observe a Bell inequality violation without the presence of loopholes \cite{Hensen2015,Lynden2015,Giustina2015}.

A midpoint among the aforementioned cases is the \textit{semi-device independent} approach based on the presence of \textit{quantum steering} \cite{Wiseman2007} to certify entanglement in the network. This is an asymmetric situation in the sense that only some of the parties in the network use trusted devices while others do not \cite{Cavalcanti2015}. Once more, trust should be understood in terms of full knowledge or characterisation of the devices used. More precisely, whenever a party's device is assumed untrusted all the analysis employed will only be based on the statistics it produces, not on its internal working. 

The steering approach is less demanding experimentally than the device-independent case, but it still presents practical interest for adversarial situations; for instance, one could think of a practical semi-device independent network composed by a single central  provider using well characterised devices, while the remaining parties, the clients, hold untrusted, \textit{i.e.} uncharacterized devices.
For these reasons the study of quantum steering has increased substantively in recent years \cite{SteeringReview}.

Methods to certify all kinds of multipartite entanglement in semi-device independent networks were presented -and experimentally demonstrated- very recently in Ref. \cite{Cavalcanti2015}. These methods rely on \textit{semi-definite programming} (SDP) techniques . They represent an important achievement for the certification of entanglement in quantum networks. In fact, these techniques certify entanglement in networks with amounts of noise that make them undetectable by the existing fully device-independent techniques \cite{Bancal2011}.

Here we apply such semi-device independent entanglement certification techniques to the three-qubit W state. Crucially, this state displays both genuine multipartite entanglement (GME) and entanglement in all of its reduced states, being then a flexible resource for the implementation of quantum networks. This is not the case for the GHZ state implemented in Ref. \cite{Cavalcanti2015}, as its reduced state turns to be separable, making therefore highly desirable to assess the certification of all types of entanglement from one single resource, like the W state. {Moreover, we show that all types of entanglement of the W state can be certified in all tripartite steering scenarios in a scheme where each party applies the same set of measurements. In this way, each party can certify all types of entanglement without the need to rely on any characterisation of the measurement devices used by the others.}

In this paper we show that by sharing a W state, every party can detect all types of entanglement without the need of trusting the measurements that the other parties are performing. This puts the W state as a promising candidate to implement asymmetric cryptographic protocols. We demonstrate the feasibility of such aforementioned techniques for entanglement certification in a proof-of-principle photonic experiment. Additionally, {we implement for the first time} the recent methods for one-sided device independent randomness certification presented in Ref. \cite{Passaro2015} to estimate the amount of randomness that can be obtained in bipartitions of the W state.

The paper is organized as follows. In \secref{sec.witnesses} we first provide a general overview on the construction of semi-device independent witnesses required to certify all types of entanglement. {Next we present theoretical results for the values achieved by such witnesses when the three-qubit W state is considered. Then, in \secref{sec.rand} we describe the scenario and methods required for one-sided device independent randomness certification,  and we apply such framework to  bipartitions of the threee-qubit W state. In \secref{sec.experimental} we present the photonic implementation and the corresponding experimental results.  Finally, in \secref{sec.conclusion} conclusions are presented.}

{\section{Entanglement detection} \label{sec.witnesses}}

In this section we first recall the methods of Ref.\cite{Cavalcanti2015} to construct multipartite semi-device independent entanglement witnesses. The intuition is the following. Assuming that the quantum state (denoted by $\rho$) distributed in the network is separable according to some particular decomposition (\textit{e.g} fully separable, separable across any bipartition, etc.) imposes constraints on the collection of all possibly observable set of post-measured states that the untrusted measurements create for the the parties holding trusted devices. From these constraints one can then determine if the original state $\rho$ could have the considered decomposition with SDP techniques in an efficient way. This SDP also provides experimentally friendly entanglement witnesses.
\
\\
\

\subsection{Semi-device independent scenarios}
\label{sec.scenarios}
\begin{figure}[h]
\centering
\includegraphics[width = .95\linewidth]{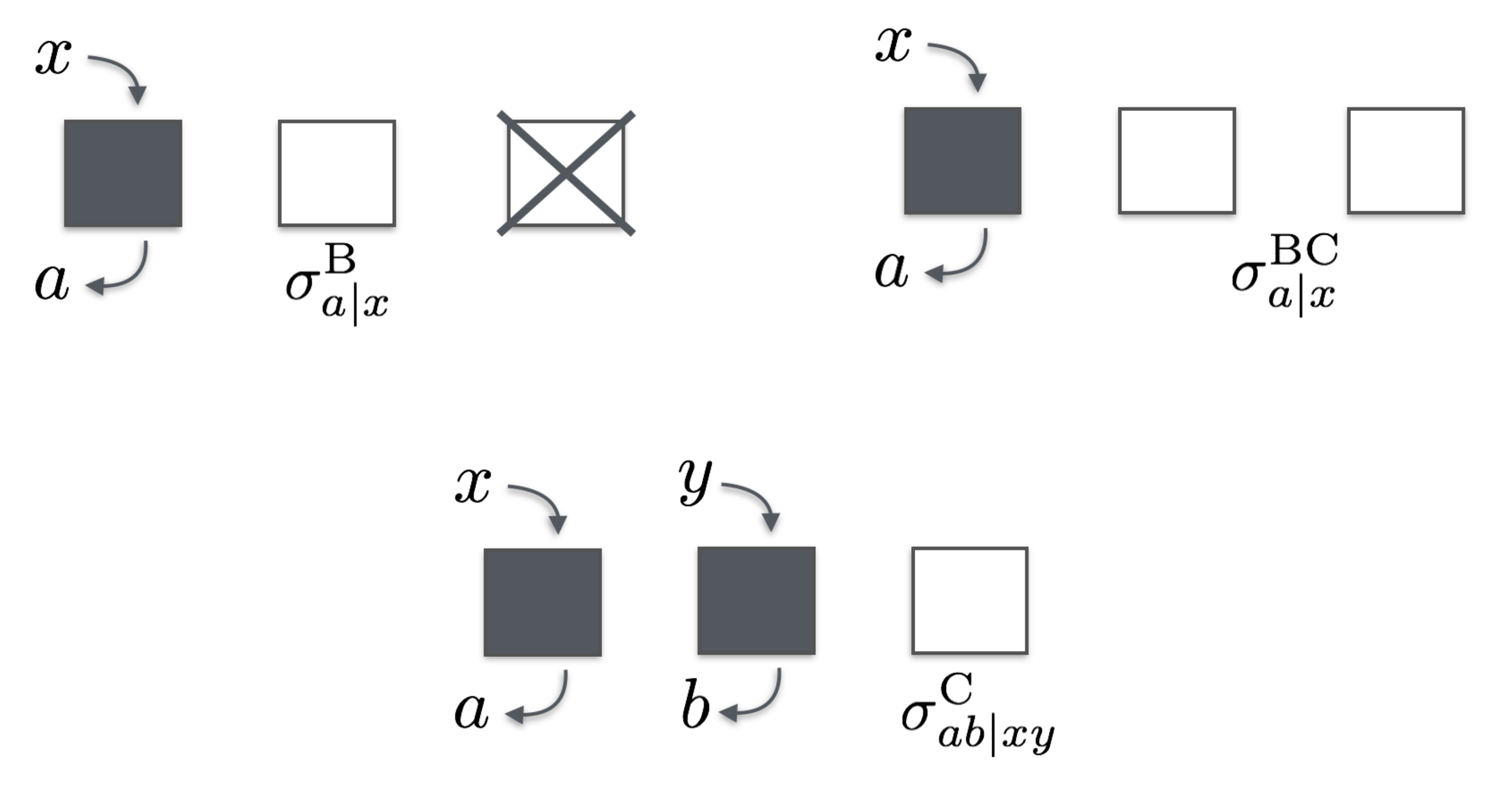}
\caption{Relevant semi-device independent networks. Trusted (untrusted) devices are represented as white (black) boxes. \textbf{Top-left:} C is discarded (crossed box) and the network consits of the reduction left to A and B. If A performs a measurement $x$ giving outcome $a$, B observes an unnormalized quantum state denoted $\sigma^{\text{B}}_{a|x}$. \textbf{Top-right:} In this case C is not discarded, and B and C observe a bipartite state $\sigma^{\text{BC}}_{a|x}$. \textbf{Bottom:} With two parties holding untrusted devices, C receives a state $\sigma^{\text{C}}_{ab|xy}$ conditioned on the statistics observed both by A and B. }
\label{fig.fig1}
\end{figure}

For illustrative purposes and simplicity we narrow our focus to the distribution of $\rho$ among three parties, A, B and C, but it is worth noting that the witness constructions that will be presented can indeed be generalized for a larger number of parties. Two semi-device independent cases arise; namely, either one or two of the parties could be holding untrusted devices in the network. In the case where one party, say A, uses untrusted devices, it will be first interesting to analyze the reduced asymmetric network that stems between A and B when discarding the third party C (top-left of \figref{fig.fig1}) . This corresponds to a bipartite steering scenario  and the witness for such case will be derived in \secref{sec.redENT}. Note that the certification of entanglement in the reduced state shared by A and B guarantees the presence of entanglement across bipartitions A:BC and AC:B, regardless of whether C trusts his measurement device or not.

A slightly more complicated task is to certify tripartite entanglement when either one or two parties hold untrusted devices (top-right and bottom of \figref{fig.fig1}, respectively). Equivalently, this task requires to decide whether the original state $\rho$ is fully separable or not, and we will construct the witness necessary to answer this question in \secref{sec.fullENT}. Next, the question will be extended for genuine multipartite entanglement (GME) in \secref{sec.GME} by asking if the state is biseparable.

\subsection{Entanglement in bipartite reductions}
\label{sec.redENT}

We start by analyzing the presence of entanglement in the reduced bipartite system left to A and B when C is disregarded (see top-left in \figref{fig.fig1}). In this case A and B are left with the quantum state $\rho^{\text{AB}}= \text{Tr}_{\text{C}} \left[\rho \right]$. The measurements performed by A on her share of $\rho^{\text{AB}}$ are untrusted and therefore they are described by unknown positive operators $M_{a|x}$ summing to identity for each $x$, where $x$ labels the measurement chosen by A and $a$ the obtained outcome. On the other hand B trusts his measurement device and can thus perform quantum tomography on his system to observe an (unnormalized) conditional state:
\begin{equation}
\sigma^{\text{B}}_{a|x} = \text{Tr}_{\text{A}}\left[ \left(M_{a|x} \otimes \mathds{1}^{\text{B}} \right) \ \rho^{\text{AB}} \right],\ \ \ \forall\ a,x.
\label{eq.sigBax}
\end{equation}
Note that the statistics observed by A can be recovered from the relation (Born rule) $p(a|x) =  \text{Tr}\left[\sigma^{\text{B}}_{a|x} \right]$, and thus the collection $\{ \sigma^{\text{B}}_{a|x}  \}_{a,x}$ known as a \textit{quantum assemblage} \cite{Wiseman2007}, contains all the information obtainable in this measurement scenario. If $\rho^{\text{AB}}$ is not entangled, it reads:
\begin{equation}
 \rho^{\text{AB}} = \sum_{\lambda} p_{\lambda}\ \rho^{\text{A}}_{\lambda}\otimes \rho^{\text{B}}_{\lambda}
 \label{eq.rhoSep}
\end{equation}
where $p_{\lambda}$ defines some probability distribution. In this case, the assemblage \eqref{eq.sigBax} takes the form:
\begin{equation}
\sigma^{\text{B}}_{a|x} = \sum_{\lambda} p(a|x\lambda)\ \rho^{\text{B}}_{\lambda},
 \label{eq.sigSep}
\end{equation}
with $p(a|x\lambda) = p_{\lambda}\text{Tr} \left[ M_{a|x}\ \rho^{\text{A}}_{\lambda}\right]$. A decomposition for $\sigma^{\text{B}}_{a|x}$ of the form \eqref{eq.sigSep} is a typical instance of an assemblage admitting a \textit{local hidden state}  (LHS) model \cite{Wiseman2007}. The collection of all such unnormalized assemblages admitting an LHS model for B forms a convex set which we denote $\Sigma_{\text{A:B}}^{\text{B}}$. Concretely:
\begin{equation}
\Sigma_{\text{A:B}}^{\text{B}} = \left\{\sigma^{\text{B}}_{a|x}\ |\ \sigma^{\text{B}}_{a|x}  = \sum_{\mu} D_{\mu}(a|x)\sigma^{\text{B}}_{\mu},\ \sigma^{\text{B}}_{\mu}\geq0\right\}
 \label{eq.sigSepSet}
\end{equation}
where we have used the fact that any probability distribution $p(a|x\lambda)$ can always be written as a convex combination of deterministic strategies $D_{\mu}(a|x)$. (Note that the constraints inside the brackets in \eqref{eq.sigSepSet} involve all values of $a$ and $x$; unless otherwise specified, all expressions with indices should be understood to hold for each value of the index.)

Crucially, imposing membership in $\Sigma_{\text{A:B}}^{\text{B}}$ involves a finite number of linear matrix inequalities and positive-semidefinite constraints for the variables $\sigma^{\text{B}}_{\mu}$ in \eqref{eq.sigSepSet}, which are all valid constraints to formulate the problem as an SDP \cite{Boyd2004}. By introducing the maximally mixed assemblage for Bob,
$\text{I}^{\text{B}}_{a|x}=\frac{1}{o_A} \text{Tr}_{\text{A}}\left[  \frac{\mathds{1}_{\text{AB}}}{d_A d_B} \right]$, where $o_A$ denotes the number outcomes of A and $d_A$ and $d_B$ denote the dimension of A's and B's respective systems, we arrive at the following SDP test for bipartite entanglement with one party holding untrusted devices:
{
\begin{equation}
\begin{array}{l}
\min\ \ r
\\
\ \text{s.t.}\ \ \ (1-r) \sigma^{\text{B}}_{a|x} + r \text{I}^{\text{B}}_{a|x} \ \in\ \Sigma_{\text{A:B}}^{\text{B}} \quad \forall\quad a,x,
\end{array}
\label{eq.entSDP}
\end{equation}
where $\sigma^{\text{B}}_{a|x}$ is the assemblage observed by B in this network. Since $\text{I}^{\text{B}}_{a|x} \in \Sigma_{\text{A:B}}^{\text{B}}$, a sufficiently small value of $r$ will always solve the second line of \eqref{eq.entSDP}, and hence this SDP is strictly feasible. The solution of \eqref{eq.entSDP},  denoted by $v^*$, quantifies how much maximally mixed noise has to be added to the assemblage such that the mixture becomes LHS: if $r^*=0$, then $\sigma^{\text{B}}_{a|x}~\in\ \Sigma_{\text{A:B}}^{\text{B}}$ and no steering can be demonstrated. On the other hand, if $r^*>0$, some amount of maximally mixed noise has to be added to the assemblage to make it LHS, so we certify entanglement in $\rho^{\text{AB}}$.}

Furthermore, from the \textit{dual} formulation of the \textit{primal} problem \eqref{eq.entSDP}, it is possible to define a set of operators $\{ F_{a|x}\}_{ax}$, which are such that the linear functional:
{
\begin{equation}
\hat{w}:\ \pi_{a|x}\ \mapsto\ \sum_{ax}\text{Tr}\left[F_{a|x}\ \pi_{a|x} \right]
\label{eq.witness1unt}
\end{equation}
provides a strictly positive value only if the assemblage $\pi_{a|x}$ demonstrates steering. Thus, $\hat{w}$ constitutes a witness for bipartite entanglement with one party holding untrusted devices. Moreover, since the primal problem is strictly feasible, \textit{strong duality} holds, and the dual and primal solutions coincide \cite{Boyd2004}. That is, $\hat{w}\left(\sigma^{\text{B}}_{a|x}\right) = r^*$, i.e. the value of the witness applied to the assemblage from which it was derived gives exactly the robustness $r^*$ of this assemblage. }

To conclude this section, note that the witness we have constructed here certifies entanglement between A and B, whether C is holding a device that is trusted or untrusted. In other words, this witness also certifies the presence of entanglement across the bipartitions A:BC and B:AC, for any level of trust that C might have on his measurement device.

\subsection{Entanglement in the full state}
\label{sec.fullENT}
We now move on to the full tripartite network, for which we wish to certify the presence of entanglement in the whole state $\rho$ distributed to A, B and C. The construction procedure is similar to the one from the previous section. If $\rho$ is fully separable, then:
\begin{equation}
 \rho = \sum_{\lambda} p_{\lambda}\ \rho^{\text{A}}_{\lambda}\otimes \rho^{\text{B}}_{\lambda} \otimes \rho^{\text{C}}_{\lambda}.
 \label{eq.rhoFullSep}
\end{equation}

We treat first the case in which only A uses an untrusted device, while B and C's devices remain trusted (see top-right of \figref{fig.fig1}). In this case, B and C are provided with the assemblage:
\begin{equation}
\sigma^{\text{BC}}_{a|x} = \text{Tr}_{\text{A}}\left[ \left(M_{a|x} \otimes \mathds{1}^{\text{B}} \otimes \mathds{1}^{\text{C}}\right)\ \rho \right],
\label{eq.sigBCax}
\end{equation}
which, after using \eqref{eq.rhoFullSep}, takes the form:
\begin{equation}
\sigma^{\text{BC}}_{a|x} = \sum_{\lambda} p(a|x\lambda)\ \rho^{\text{B}}_{\lambda}\otimes \rho^{\text{C}}_{\lambda}.
 \label{eq.sigFullSep}
\end{equation}
A decomposition of $\sigma^{\text{BC}}_{a|x}$ of the form \eqref{eq.sigFullSep} can readily be seen to be similar to the one in \eqref{eq.sigSep}, with the only difference that now the bipartite states of B and C conditioned on $a$ and $x$ are separable. This last requirement (separability) cannot, in general, be translated to a finite number of linear matrix inequalities and positive constraints as before, because the set of separable states has a complicated structure. However, separability can be relaxed to \textit{positivity under partial transposition} \cite{Horodecki2009}, which is now a valid SDP constraint -and equivalent to separability whenever the product of the dimensions of B and C satisfies $d_Bd_C\leq 6$. Therefore, we define the relaxed set:
\begin{equation}
\begin{array}{r}
\Sigma_{\text{A:B:C}}^{\text{BC}} = \left\{\sigma^{\text{BC}}_{a|x}\ |\ \sigma^{\text{BC}}_{a|x}  = \sum_{\mu} D_{\mu}(a|x)\sigma^{\text{BC}}_{\mu},\right.
\\
\ \left. \sigma^{\text{BC}}_{\mu}\geq0,\ \left(\sigma^{\text{BC}}_{\mu}\right)^{T_{\text{B}}} \geq0 \right\}
\end{array}
 \label{eq.sigFullSepSet}
\end{equation}
where $T_{\text{B}}$ denotes the partial transposition operation with respect to system B.  In the same fashion as in \secref{sec.redENT}, with the help of the maximally mixed assemblage for B and C, namely $\text{I}^{\text{BC}}_{a|x}=\frac{1}{o_A} \text{Tr}_{\text{A}}\left[  \frac{\mathds{1}_{\text{ABC}}}{d_A d_B d_C} \right]$, we obtain the corresponding SDP test for tripartite entanglement with one party holding untrusted devices:
{
\begin{equation}
\begin{array}{l}
\min\ \ r
\\
\ \text{s.t.}\ \ \ (1-r)\sigma^{\text{BC}}_{a|x}+r\text{I}^{\text{BC}}_{a|x} \ \in\ \Sigma_{\text{A:B:C}}^{\text{BC}}.
\end{array}
\label{eq.FullentSDP}
\end{equation}
}
Here, again, duality theory allows one to retrieve operators $\{ F_{a|x}\}_{ax}$ defining a new witness $\hat{w}$ with the exact same structure \footnote{The only difference is that the witness now is a functional acting on the space of assemblages of B and C.} as in \eqref{eq.witness1unt}, and such that  $\hat{w}\left(\sigma^{\text{BC}}_{a|x}\right)<0$ guarantees that $\rho$ is tripartite entangled. Finally, note that whenever $d_Bd_C > 6$, there exist entangled steerable states that remain undetected by our witness construction, since the set of unsteerable states does not coincide with  $\Sigma_{\text{A:B:C}}^{\text{BC}}$ in this case. But since $\Sigma_{\text{A:B:C}}^{\text{BC}}$ does contain all separable states, a {positive} value of the witness guarantees that the state is, unequivocally, entangled.

In the case of two parties holding untrusted devices (say A and B, as shown at the bottom of \figref{fig.fig1}), C observes the assemblage:
\begin{equation}
\sigma^{\text{C}}_{ab|xy} = \text{Tr}_{\text{AB}}\left[\left( M_{a|x} \otimes M_{b|y} \otimes \mathds{1}^{\text{C}}\right)\ \rho \right],
\label{eq.sigCabxy}
\end{equation}
which, after replacing $\rho$ with its separable form \eqref{eq.rhoFullSep}, gives:
\begin{equation}
\sigma^{\text{C}}_{ab|xy} = \sum_{\lambda} p(ab|xy\lambda)\ \rho^{\text{C}}_{\lambda}.
 \label{eq.sigFullSep2unt}
\end{equation}
Since $p(ab|xy\lambda)$ arises from local measurements of a separable state it must be \textit{local} \cite{BrunnerCavalcantiPironioScaraniWehner2013}, and can therefore be written as a convex combination of products of determinstic strategies for A and B. Thus, the relevant set of unnormalized assemblages for tripartite entanglement with two parties using untrusted measurements is:
\begin{equation}
\begin{array}{l}
\Sigma_{\text{A:B:C}}^{\text{C}} = \left\{\sigma^{\text{C}}_{ab|xy}\ | \right.
\\
\ \ \ \ \ \left.\sigma^{\text{C}}_{ab|xy}  = \sum_{\mu\nu} D_{\mu}(a|x)D_{\nu}(b|y)\sigma^{\text{C}}_{\mu\nu},\ \sigma^{\text{C}}_{\mu\nu}\geq0\right\}
\end{array}
 \label{eq.sigFullSepSet2unt}
\end{equation}
and since membership in $\Sigma_{\text{A:B:C}}^{\text{C}}$ involves valid SDP constraints, we obtain the corresponding SDP test:
{
\begin{equation}
\begin{array}{l}
\min\ \ r
\\
\ \text{s.t.}\ \ \ (1-r)\sigma^{\text{C}}_{ab|xy}+r\text{I}^{\text{C}}_{ab|xy} \ \in\ \Sigma_{\text{A:B:C}}^{\text{C}},
\end{array}
\label{eq.FullentSDP2unt}
\end{equation}
}
where $\text{I}^{\text{C}}_{ab|xy}=\frac{1}{o_A o_B} \text{Tr}_{\text{AB}}\left[  \frac{\mathds{1}_{\text{ABC}}}{d_A d_B d_C} \right]$ is the maximally mixed assemblage for C. The set of dual variables $\{F_{ab|xy} \}_{abxy}$ of program \eqref{eq.FullentSDP2unt} enable the construction of the witness:
{
\begin{equation}
\hat{w}:\ \pi_{ab|xy}\ \mapsto\ \sum_{abxy}\text{Tr}\left[F_{ab|xy}\ \pi_{ab|xy} \right],
\label{eq.witness2unt}
\end{equation}
which provides a strictly positive value if and only if $ \pi^{\text{C}}_{ab|xy}$ demonstrates steering. Once more, $\hat{w}(\sigma_{ab|xy}^\text{C})=r^*$.}

\subsection{Genuine Multipartite Entanglement}
\label{sec.GME}
In this section we move to the more delicate task of certifying genuine tripartite entanglement in the state $\rho$ shared by A, B and C. If $\rho$ is not genuinely tripartite entangled, it is biseparable and has the form:
\begin{equation}
 \rho = \sum_{\lambda} \left( p_{\lambda}\rho^{\text{A}}_{\lambda}\otimes \rho^{\text{BC}}_{\lambda} + q_{\lambda}\rho^{\text{B}}_{\lambda}\otimes \rho^{\text{AC}}_{\lambda} + r_{\lambda}\rho^{\text{C}}_{\lambda}\otimes \rho^{\text{AB}}_{\lambda} \right)
 \label{eq.rhoBiSep}
\end{equation}
where $p_{\lambda}$, $q_{\lambda}$ and $r_{\lambda}$ are unnormalised probability distributions. For the case of one party using an untrusted device (top-right of \figref{fig.fig1}), using \eqref{eq.rhoBiSep} in \eqref{eq.sigBCax} leads to:
\begin{equation}
 \sigma_{a|x}^{\text{BC}}= \sum_{\lambda} p_{\lambda}p(a|x\lambda)\rho^{\text{BC}}_{\lambda} + q_{\lambda}\rho^{\text{B}}_{\lambda}\otimes \sigma^{\text{C}}_{a|x\lambda} + r_{\lambda} \sigma^{\text{B}}_{a|x\lambda}\otimes\rho^{\text{C}}_{\lambda}
 \label{eq.sigGME1unt}
\end{equation}
The first summation term in \eqref{eq.sigGME1unt} has the same structure as the decomposition in \eqref{eq.sigFullSep}, with the only difference that now $\rho^{\text{BC}}_{\lambda}$ may even be entangled. Thus, the relevant set defined for this first decomposition term is:
\begin{equation}
\Sigma_{\text{A:BC}}^{\text{BC}} = \left\{\sigma^{\text{BC}}_{a|x}\ |\ \sigma^{\text{BC}}_{a|x}  = \sum_{\mu} D_{\mu}(a|x)\sigma^{\text{BC}}_{\mu},\sigma^{\text{BC}}_{\mu}\geq0\right\}
 \label{eq.sigGMEsetABC}
\end{equation}
which clearly involves valid SDP constraints only. The second summation term in \eqref{eq.sigGME1unt} has two main features: \textit{(i)} it does not demonstrate steering from A to B. This implies that tracing out C will leave B with an LHS assemblage with respect to A. \textit{(ii)} it is separable across B:C. The first feature, having an LHS model, is readily translated to SDP constraints, as we have seen in the previous sections. The second one, separability, has to be relaxed to positivity under partial transposition, as explained in \secref{sec.fullENT}. Thus, the relaxed set $\Sigma_{\text{B:AC}}^{\text{BC}}$ corresponding to the second decomposition term is given by:
\begin{equation}
\begin{array}{r}
\Sigma_{\text{B:AC}}^{\text{BC}} = \left\{\sigma^{\text{BC}}_{a|x}\ |\ \text{Tr}_{\text{C}}\left[\sigma^{\text{BC}}_{a|x}\right]  = \sum_{\mu} D_{\mu}(a|x)\sigma^{\text{B}}_{\mu},\right.
\\
\ \left. \sigma^{\text{B}}_{\mu}\geq0,\ \left(\sigma^{\text{BC}}_{a|x}\right)^{T_{\text{B}}} \geq0 \right\}.
\end{array}
\label{eq.sigGMEsetBAC}
\end{equation}
The third summation term of \eqref{eq.sigGME1unt} is identical to the second term, except that the roles of B and C are interchanged. From the previous claims we can now write explicitly the SDP test for genuine tripartite entanglement with one party holding untrusted devices:
{
\begin{equation}
\begin{aligned}
& \min
& & r \\
& \ \text{s.t.} & & (1-r)  \sigma^{\text{BC}}_{a|x}+r\text{I}^{\text{BC}}_{a|x} = \gamma_{a|x}^{\text{A:BC}} + \gamma_{a|x}^{\text{B:AC}} + \gamma_{a|x}^{\text{C:AB}} \\
& & &  \text{\small{$\gamma_{a|x}^{\text{A:BC}} \in\ \Sigma_{\text{A:BC}}^{\text{BC}},\ \gamma_{a|x}^{\text{B:AC}} \in\ \Sigma_{\text{B:AC}}^{\text{BC}},\ \gamma_{a|x}^{\text{C:AB}} \in\ \Sigma_{\text{C:AB}}^{\text{BC}}$}}\\
\end{aligned}
\label{eq.GME1untSDP}
\end{equation}
and the dual formulation of this problem provides a witness $\hat{w}$, defined as in \eqref{eq.witness1unt}, such that $\hat{w}\left( \sigma^{\text{BC}}_{a|x} \right) > 0$ certifies that $\rho$ is genuinely tripartite entangled.}

For the case of two parties using untrusted devices (bottom of \figref{fig.fig1}), substituting \eqref{eq.rhoBiSep} in \eqref{eq.sigCabxy} gives:
\begin{equation}
\begin{array}{c}
 \sigma_{ab|xy}^{\text{C}}= \sum_{\lambda} \left( p_{\lambda}p(a|x\lambda)\sigma^{\text{C}}_{b|y\lambda} + q_{\lambda}p(b|y\lambda)\sigma^{\text{C}}_{a|x\lambda} +\right.
 \\
 \left. r_{\lambda}p(ab|xy\lambda) \sigma^{\text{C}}_{\lambda} \right).
 \label{eq.sigGME2unt}
\end{array}
\end{equation}
The first summation term of the decomposition obtained in \eqref{eq.sigGME2unt} has only one main feature, that it may contain steering from B to C, but never from A to C. Hence, it defines the set $\Sigma_{\text{A:BC}}^{\text{C}}$ as:
\begin{equation}
\begin{array}{l}
\Sigma_{\text{A:BC}}^{\text{C}} = \left\{\sigma^{\text{C}}_{ab|xy}\ | \right.
\\
\ \ \ \ \ \left.\sigma^{\text{C}}_{ab|xy}  = \sum_{\mu} D_{\mu}(a|x)\sigma^{\text{C}}_{b|y\mu},\ \sigma^{\text{C}}_{b|y\mu}\geq0\right\}.
\end{array}
 \label{eq.sigBisepSet2unt}
\end{equation}
The second summation term appearing in \eqref{eq.sigGME2unt} has the same structure as the first one, and thus the set $\Sigma_{\text{B:AC}}^{\text{C}}$ is straightforwardly defined in the same manner as $\Sigma_{\text{A:BC}}^{\text{C}}$ in \eqref{eq.sigBisepSet2unt}. The third term in \eqref{eq.sigGME2unt}, on the other hand, has two features: \textit{(i)} it corresponds to an LHS assemblage, and \textit{(ii)} the probability distribution $p(ab|xy\lambda)$ is \textit{quantum}, in the sense that it arises from local measurements performed on (possibly entangled) states $\rho^{\text{AB}}_\lambda$. Deciding if a probability distribution is quantum is not straightforward because the quantum set $Q$ lacks of a precise characterization. Nevertheless, it is possible to construct a hierarchy $\{Q^{(k)}\}_k$ of well characterized SDP relaxations of the set $Q$, the so-called Navascu\'es-Pironio-Ac\'in (NPA) hierarchy \cite{Navascues2007,Navascues2008}. The NPA hierarchy, originally introduced for probability distributions, can be generalized to assemblages, in the sense that one can define the $k$-th relaxed set $Q^{(k)}_{\text{C}}$ of assemblages for C  encoding all the constraints on orthogonality and commutativity for the measurement operators of A and B (see \cite{Cavalcanti2015} for more details). Given the above, the relaxed set $\Sigma_{\text{C:AB}}^{\text{C}}$ is:
\begin{equation}
\begin{array}{l}
\Sigma_{\text{C:AB}}^{\text{C}(k)} = \left\{\sigma^{\text{C}}_{ab|xy}\ | \right.
\\
\ \ \ \ \ \left.\sigma^{\text{C}}_{ab|xy}  = \sum_{\mu} D_{\mu}^{NS}(a|x)\sigma^{\text{C}}_{\mu},\ \sigma^{\text{C}}_{ab|xy}\in Q^{(k)}_{\text{C}} \right\},
\end{array}
 \label{eq.sigBisepSetBis2unt}
\end{equation}
where we had to use the finite set of non-signalling strategies $\{D_{\mu}^{NS}(a|x)\}_{\mu}$  to decompose the quantum probability distribution $p(ab|xy\lambda)$ in terms of a finite number of distributions \cite{BrunnerCavalcantiPironioScaraniWehner2013} (deterministic strategies are insufficient here because of the existence of quantum correlations lying outside the local set). The SDP test for genuine tripartite entanglement where two parties use untrusted devices is, therefore:
{
\begin{equation}
\begin{aligned}
& \min
& & r \\
& \ \text{s.t.} & &  (1-r) \sigma^{\text{C}}_{ab|xy}+r\text{I}^{\text{C}}_{ab|xy} = \gamma_{ab|xy}^{\text{A:BC}} + \gamma_{ab|xy}^{\text{B:AC}} + \gamma_{ab|xy}^{\text{C:AB}} \\
& & &  \text{\small{$\gamma_{ab|xy}^{\text{A:BC}} \in\ \Sigma_{\text{A:BC}}^{\text{C}},\ \gamma_{ab|xy}^{\text{B:AC}} \in\ \Sigma_{\text{B:AC}}^{\text{C}},\ \gamma_{ab|xy}^{\text{C:AB}} \in\ \Sigma_{\text{C:AB}}^{\text{C}(k)}$}}\\
\end{aligned}
\label{eq.GME2untSDP}
\end{equation}
}
and the corresponding semi-device independent witness $\hat{w}$ is obtained from duality theory and has the same structure as \eqref{eq.witness2unt}.
\
\\
\

{\subsection{Multipartite steering of the W state} \label{sec.theoretical}}

Here we provide numerical values for the three-qubit W state 
\begin{equation}
\ket{W} = \frac{1}{\sqrt{3}}(\ket{001}+\ket{010}+\ket{100})
\end{equation} 
and discuss the fact that parties A, B, and C can check for all kinds of entanglement without trusting the devices of the others. In all that follows we consider that the measurements performed by all trusted boxes are Pauli observables, namely, $\hat{X}$, $\hat{Y}$ and $\hat{Z}$. {Our theoretical findings regarding the witness values $r$ are summarized in \tabref{tab.1}}.

Since the W state is symmetric, all reductions are equivalent regardless of the party that is discarded. Specifically, $\rho_{\text{red}} =2/3\ket{\psi^+}\bra{\psi^+}+1/3\ket{00}\bra{00}$, where $\ket{\psi^+}$ denotes the two-qubit maximally entangled state $\ket{\psi^+}= 1/\sqrt{2}(\ket{01}+\ket{10})$. The reduced state turns out to be steerable with a theoretical violation of {$r^*=0.11$}. This value is relatively small because of the detrimental contribution of the separable state $\ket{00}$. Such violation not only guarantees the presence of entanglement in the reduced state, but it also certifies the presence of entanglement across any bipartition of the tripartite network, regardless of whether the third party (the discarded one) is using a trusted device or not, as explained in the first paragraph of \secref{sec.scenarios}. {We also notice that  it is still an open question whether the reduced state of the W state can violate any Bell inequality \cite{Sohbi2015,Barnea2015}, while here we show that it does present steering.}

For the tripartite W state, we observe the presence of entanglement and GME both in the ``one untrusted'' scenario and in the ``two untrusted'' scenario as well (see lines 2-5 of \tabref{tab.1}). Note that the violations for the ``one untrusted'' scenario are always better than for the ``two untrusted'' scenario, because in the former case there is more useful information available (about the state) than in the latter case. The values for tripartite entanglement are also always better than the values obtained for GME, as the presence of the latter implies the presence of the former, but the converse is not true in general; that is, a state might be entangled without being GME.

{\section{One-sided device independent randomness certification}\label{sec.rand}}

\subsection{Scenario and guessing probability}
\label{sec.randscenario}

One of the most interesting applications following the semi-device-independent quantum information approach is the \textit{semi-device independent random number generation} \cite{Law2014, Passaro2015}. To carry out this task, one must bound the predictability of the outcomes of the black boxes (parties holding untrusted devices), only from the observation of a certain assemblage by the parties holding trusted devices in the network. Just as in the previous sections, the black boxes might be provided with any \textit{a priori} unknown quantum state $\rho$, but now $\rho$ may even be correlated with some other quantum system $\rho^{\text{ABCE}}$ in the possession of a malicious eavesdropper E, such that $\rho=\text{Tr}_{\text{E}}\left[ \rho^{\text{ABCE}} \right]$. Furthermore, semi-device independent random number generation is appealing because it enables a corresponding cryptographic task, namely, \textit{semi-device independent quantum key distribution} \cite{Branciard2012}.

\begin{figure}[h]
\centering
\includegraphics[width = .5\linewidth]{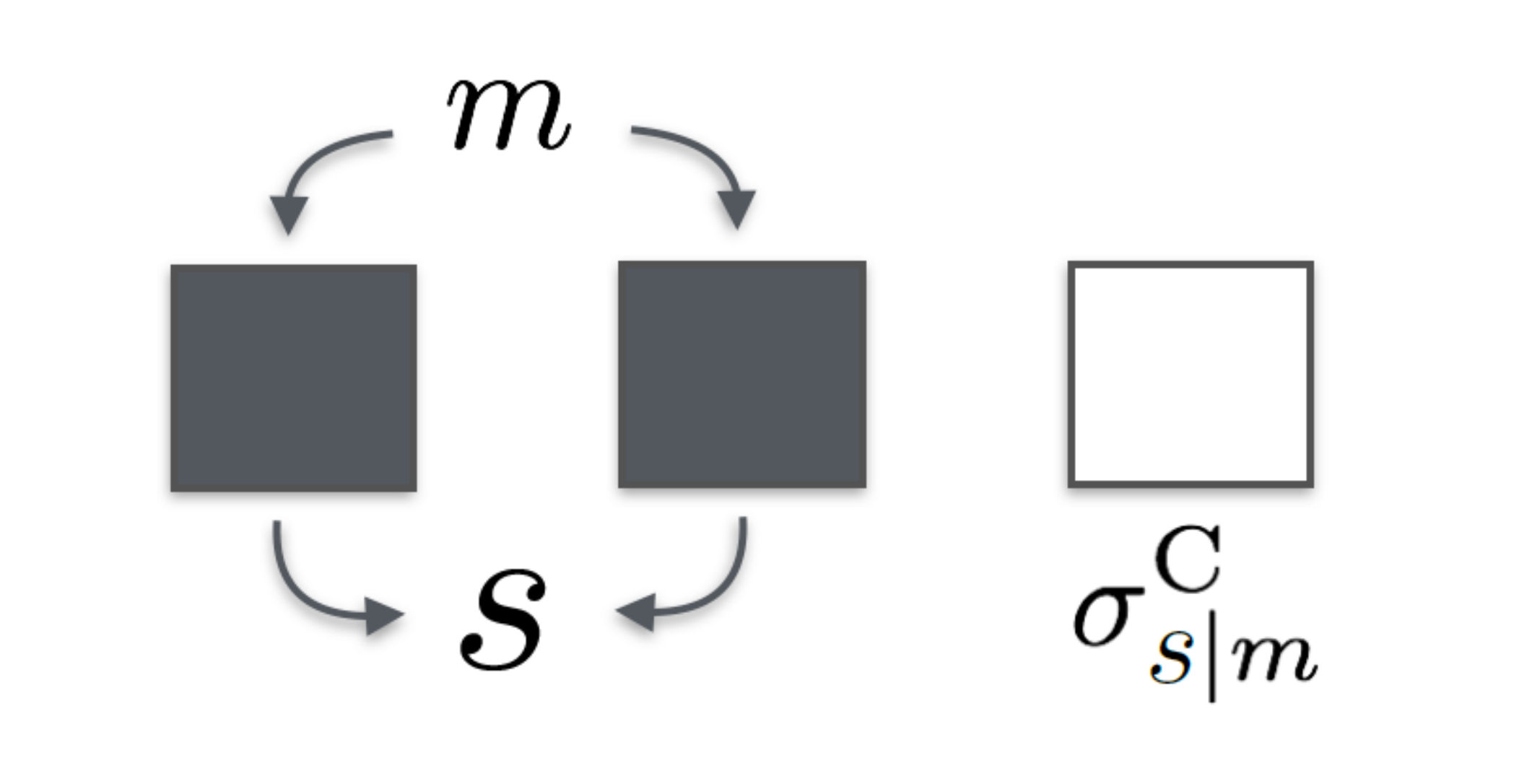}
\caption{\textbf{One-sided randomness certification scenario.} A and B are thought as a single party holding two untrusted boxes, performing at each round a measurement $m$ and obtaining some result $s$. This scenario allows us to analyze the amount of randomness of the result $s$ when $\sigma^{\text{C}}_{s|m}$ is observed by C.}
\label{fig.fig2}
\end{figure}

As an interesting example, here we implement the methods of Ref. \cite{Passaro2015} to certify the optimal amount of one-sided randomness present in the string of outcomes of two black boxes, when the third one remains trusted. Since these methods can only be applied on a bipartite scenario, we shall consider that the two black boxes are held by a single party performing measurements labeled $m=(x,y)$ and obtaining outcomes labeled $s=(a,b)$, as illustrated in \figref{fig.fig2}.

The predictability of the outcome $s$ when a given measurement $m^*$ is chosen is quantified by the \textit{guessing probability} $G_{\sigma}(m^*)$: the probability that E guesses correctly the value of $s$, optimized over all of E's possible strategies, and conditioned on the observation of the assemblage $\sigma^\text{C}_{s|m}$ by the party C \cite{Pironio2014, Passaro2015}. Formally:
\begin{equation}
\begin{aligned}
& G_{\sigma}(m^*)\ =\ \text{max} \ \sum_{e}\text{Tr}\left[ \sigma_{s=e|m^*}^e \right]\\
&\ \ \ \ \ \ \ \ \ \ \ \ \ \ \ \ \ \text{s.t.}\ \ \ \text{\small{$\sum_{e}\sigma^e = \sigma$}}\\
& \ \ \ \ \ \ \ \ \ \ \ \ \ \ \ \  \ \ \ \ \ \ \ \ \text{\small{$\sum_{s}\sigma_{s|m}^e = \sum_{s}\sigma_{s|m'}^e\ \ \forall e,m,m'$}}\\
& \ \ \ \ \ \ \ \ \ \ \ \ \ \ \ \  \ \ \ \ \ \ \ \ \ \text{\small{$\sigma^e_{s|m}\geq 0 \ \ \ \forall s,m,e$}}
\end{aligned}
\label{eq.Randomness}
\end{equation}
where we used $\text{Tr}\left[ \sigma_{s|m^*}^e \right] = p(e)p(s|e,m^*)$ to re-express the objective function in the first line of \eqref{eq.Randomness}. Intuitively, Eve may prepare any convex combination of the unnormalized assemblages  $\{\sigma_{s|m}^e\}_e$, which are such that, whenever Eve obtains the result $e=s$ after measuring her system, she then guesses that the outcome of the black boxes was $s$. Thus, Eve needs as many preparations  $\{\sigma_{s|m}^e\}_e$ as possible values that $s$ can take, and $\sigma_{s=e|m}^e$ would be the assemblage obtained if the information of Eve's outcome $e$ were available. The first constraint in \eqref{eq.Randomness} translates the fact that Eve has to reproduce on average the observed assemblage, since otherwise her attack would be detected by the party holding the trusted device. The second and third constraints, non-signalling and positivity respectively, guarantee that the assemblages prepared by Eve stem from quantum theory. In fact, any bipartite non-signalling assemblage admits a quantum realization \cite{Hughston1993,Sainz2015}.

Once again, duality theory allows us here -from the dual formulation of \eqref{eq.Randomness}- to obtain a steering inequality (\textit{i.e.} a linear functional $\hat{w}$) acting on the set of assemblages of C, having the same structure as \eqref{eq.witness2unt}, and such that $\hat{w}\left( \sigma \right) = G_{\sigma}(m^*)$. Finally, assuming Independent and Identically Distributed (IID) rounds in the experiment, the amount of genuine random bits certified is given by $R = -\log_2\left( G_{\sigma}(m^*) \right)$, the \textit{min-entropy} of the semi-device independent guessing probability \cite{Pironio2014}. Note that, when $G_{\sigma}(m^*) = 1$ no random bits can be certified ($R=0$), as E guesses with probability 1 the outcome of the boxes and thus there is no unpredictability. On the other hand, any value $G_{\sigma}(m^*) < 1$, guarantees unpredictability and hence provides a strictly positive amount of randomness $R>0$.

{ \subsection{One-sided device independent randomness certification from bipartitions of the W state} }
\label{sec.resultsrandth}

We first analyzed the reduced state $\rho_{\text{red}}$ of the three-qubit W state. Though we found that the state is steerable (as shown in the first line of \tabref{tab.1}), it is not possible to extract one-sided genuine randomness from such reduced state with the measurements considered here. On the other hand, one-sided randomness certification turns out to be appealing when bipartitions of the tripartite W state are considered. In fact, when two of the boxes are seen as a single untrusted measurement $m=(x,y)$ performed on some unknown quantum state, while the other qubit, C, uses trusted measurements (see \figref{fig.fig2}), we manage to certify $-\text{log}_2(1/3) \approx 1.58$ random bits from the outcome $s=00,01,10,11$ of the measurement $m^*$ corresponding to the observables $\hat{X}$ and $\hat{Z}$ acting on the two-qubit subspace reduction of the W state.

The fact that no one-sided genuine randomness could be extracted from the reduced state of the three-qubit W state is unfortunate but interesting. This is analogous to a similar phenomenon, known as \textit{bound randomness} \cite{Acin2015b}, which arises in the fully device independent scenario where the two parties hold untrusted devices. More precisely, in the fully device independent scenario, bound randomness arises in nonlocal correlations for which a non-signalling eavesdropper can find out a posteriori the result of any implemented measurement. Thus, the fact that any reduction of the W state is steerable but does not allow for one-sided randomness certification, may tentatively be refered to as \textit{one-sided bound randomness}: a form of steerable correlations for which an eavesdropper can predict the result of any of the measurements performed on the untrusted side.
\
\\
\

\
\\
\

\section{Experimental Results}
\label{sec.experimental}
\subsection{Setup}
\label{sec.setup}
\
\\
\

\begin{figure}
\centering
\includegraphics[width = \linewidth]{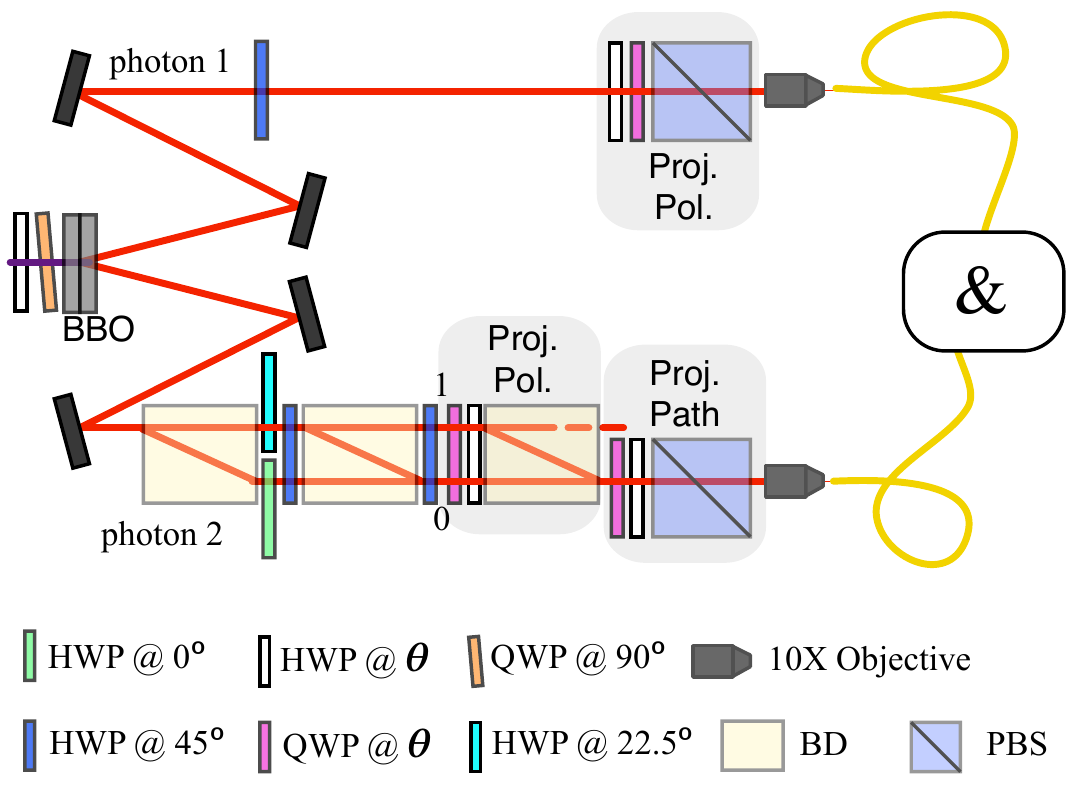}
\caption{{Experimental setup.}  Polarization-entangled photons are produced using SPDC.  A third qubit is encoded in the path degree of freedom of photon 2.  An interferometer consisting of beam displacers is used to produce a three-qubit W state. Projective measurements on the polarization (Proj. Pol.) and Path (Proj. Path) measurements are performed using wave plates and polarizers.  Additional details are provided in the main text.}
\label{fig:setup}
\end{figure}

To demonstrate the practical utility of the theoretical results presented in sections \ref{sec.witnesses} and \ref{sec.theoretical}, we produced a three-qubit W state using photon pairs produced by Spontaneous Parametric Down Conversion (SPDC).    Figure \ref{fig:setup} shows the experimental setup.  Two 1mm thick type-I non-linear BBO crystals with optical axes oriented perpendicularly were pumped with a 325nm continuous-wave He-Cd laser, producing degenerate photon pairs centered around 650nm.  Using an additional half-wave plate in the path of photon 1, the crossed-crystal arrangement produces two polarization entangled photons in the target state \cite{kwiat99}:
\begin{equation}
\ket{\psi} = \cos \theta \ket{VH}_{12} + e^{i \varphi} \sin\theta \ket{HV}_{12}. 
\label{eq:polstate}
\end{equation}  

Qubits $B$ and $C$ were encoded in the polarization of the photons 1 and 2, respectively. Qubit $A$ was encoded in the path of photon 2.  Initially, qubit $A$ is in the state $\ket{0}_A$. To produce the W state, we entangle the path and polarization degrees of freedom (DOF) of photon 2 using a polarization-dependent Mach-Zehnder interferometer composed of two beam displacers (BDs) and several half-wave plates (HWPs), as  described in more detail in Refs. \cite{farias12b,aguilar14c}.  The angles of the HWPs are shown in Figure \ref{fig:setup}. We label the input and output paths (0 and 1 in the figure) such that when the polarization state is $\ket{H}_{C}$, the output state is $\ket{0H}_{AC}$. For input vertical polarization, the interferometer implements the transformation   
\begin{equation}
\ket{0V}_{AC} \longrightarrow \frac{1}{\sqrt{2}} \ket{0V}_{AC} + \frac{1}{\sqrt{2}} \ket{1H}_{AC}.
\label{eq:interf}
\end{equation} 
By controlling the polarization of the pump laser \cite{kwiat99}, the initial polarization entangled state was prepared with $\cos\theta={1}/{\sqrt{3}}$ and $\varphi=0$. Renaming the polarization state $\ket{H} \rightarrow \ket{0}$ and $\ket{V} \rightarrow \ket{1}$, our setup produces a  three-qubit state that is ideally a $W$ state \cite{farias12b}. {In \appref{app.exp} we provide details on the stability of the setup, the characterisation of the W state and of its reconstructed density matrix.} 
\
\\
\

\
\\
\

\
\\
\

A set of 216 joint projective measurements in the $\hat{X}$, $\hat{Y}$ and $\hat{Z}$ Pauli basis was performed on all three qubits, which allowed us to evaluate the SDP tests developed above.  To perform projective measurements on qubit $B$ (polarization of photon 1), the usual system consisting of a quarter-wave plate (QWP), HWP and a polarizing beam splitter (PBS) is used. For projective measurements on qubit $C$, a QWP, HWP and BD are used.  This measurement system works in much the same way as that of qubit $B$, however, after the projection on a given polarization state, the BD maps the state of the path DOF at its input into the polarization DOF at its output.  In this fashion, the state describing the path DOF, which is now encoded in the polarization DOF, can be measured using the same arrangement as in photon 1.   The photons were spectrally filtered with 3nm FWHM bandwidth filters centered at 650nm (not shown in the figure), coupled into single-mode optical fibers using $10 \times$ microscope objectives and registered with single photon detectors and coincidence electronics (the "\&" box in Fig. \ref{fig:setup}).  

\begin{table*}[t]\centering
\ra{1.3}
\begin{tabular}{r c@{\hskip 0.1cm} c c@{\hskip 0.1cm} l l}\toprule[1pt]

\phantom{abc} &\multicolumn{2}{c}{\parbox[t]{2.7cm}{\textbf{Theory}}} &\multicolumn{2}{c}{\parbox[t]{2cm}{\textbf{Experiment}}}\\
 \rule{0pt}{4ex}
  Ent. in reductions\   &\multicolumn{2}{c}{$0.1112$}  &\multicolumn{2}{c}{$0.07\pm0.01$} \\
  Ent. \textit{(1 untrusted)}   &\multicolumn{2}{c}{$0.7297$}  &\multicolumn{2}{c}{$0.77\pm0.01$} \\
  Ent. \textit{(2 untrusted)}   &\multicolumn{2}{c}{$0.5355$}  &\multicolumn{2}{c}{$0.500\pm0.008$}  \\
  GME \textit{(1 untrusted)}   &\multicolumn{2}{c}{$0.4581$}  &\multicolumn{2}{c}{$0.41\pm0.01$}  \\
  GME \textit{(2 untrusted)}   &\multicolumn{2}{c}{$0.3244$}  &\multicolumn{2}{c}{$0.32\pm0.01$}  \\
\bottomrule[1pt]
\end{tabular}
\caption{\textbf{Witness values $r^*$ from the three qubit W state}. For each of the quantum information tasks appearing in the first column, the corresponding witness is constructed as explained in \secref{sec.witnesses} and in \secref{sec.considerations}. A strictly positive value of the semi-device independent witness certifies the presence of: entanglement in the reduced state (first row), entanglement in the full tripartite state (second and third rows), genuine multipartite entanglement (fourth and fifth rows). See main text for details.}
\label{tab.1}
\end{table*}

\subsection{Practical considerations}
\label{sec.considerations}
The methods described in \secref{sec.witnesses} were designed to detect entanglement and certify randomness from an observed \textit{physical} assemblage $\sigma^{\text{phys}}$. However, due to the unavoidable problem of finite statistics in any experiment, the assemblage that is experimentally observed $\sigma^{\text{exp}}$ does not satisfy the non-signalling property. To overcome this problem we took the following steps. First, we construct a physical assemblage that approximates the experimental data. This step is done with a least-squares optimization, an SDP program that minimizes the distance from the experimental assemblage to the set of physical assemblages bounded by the non-signalling constraints. The second step consists of using the constructed physical assemblage to obtain the desired witness $\hat{w}^{\text{phys}}$, following the SDP techniques for entanglement detection. The last step is to apply the derived witness, which is simply a linear functional, to the experimental assemblage to show the presence of entanglement/randomness in the network.

{ \subsection{Experimental multipartite entanglement detection in the steering scenario} }
\label{sec.results}

Our experimental results are summarised in \tabref{tab.1}. The error bars were calculated by performing Monte Carlo simulation (494 rounds) assuming Poissonian coincidence counting statistics of our measurement results. Experimentally, the reduced state is not entirely symmetric because of imperfections in the optical setup, such as temporal and spatial mode mismatch in the interferometer. Thus, we analyzed all reductions and found that the highest violation of $0.07\pm0.01$ is obtained when discarding party $B$, corresponding to the polarization of photon 1.  This is due to the fact that the entanglement that survives, between qubits $A$ and $C$, is encoded entirely in the polarization and spatial qubits of photon 2.

As far as the experimental certification of tripartite entanglement and genuine multipartite entanglement (GME) are concerned, the corresponding observed witness values are shown in lines 2-3 and 4-5 of \tabref{tab.1}, respectively. One obtains a strictly positive value (violation) for these two types of tripartite entanglement, both in the ``one untrusted'' and in the ``two untrusted'' scenarios. The experimental witness values are close to the theoretical ones, although these do not always fall within the error margins obtained. This is expected since the experimental state is not perfectly pure (see \appref{app.exp}).  The case where the measured value agrees with the theory within the error interval corresponds to the situation where the correlations between two internal degrees of freedom of the same photon (path and polarization) are the most relevant. In this special case the purity of the reduced state can be very high experimentally.  Even with these small discrepancies between theory and experiment, we successfully certify the presence of entanglement and GME in the considered semi-device independent networks. {Please note that, for completeness the reader may find the numerical values for all steering inequalities described in this work at the Git online repository: github.com/mattarcon2tes/Steering}.

{ \subsection{Experimental one-sided device independent randomness certification} }
\label{sec.resultsrandexp}
The scenario presented in \figref{fig.fig2} turns out to be relevant and well suited for our experimental implementation of the photonic W state (see \secref{sec.experimental}), as a physical bipartition of the state stems between the two photons produced. In this sense, it is natural to consider the photon encoding both polarization and path qubits as a single  party, and analyze the amount of randomness of the outcomes $s=(a,b)$ retrieved when untrusted measurements are performed on such physical part of the system.

First of all, we checked that the experimental data of each of the the reduced states does not reveal any amount of one-sided randomness, as predicted by the theory. Subsequently, we managed to certify $0.26\pm0.04$ bits from the bipartitions of the W state. This value falls far from the theoretical value of $-\log_2(2/3)\approx 1.58$ bits. This discrepancy is due to the fact that the amount of randomness is extremely sensitive to the visibility of the pure $W$ state with respect to white noise. For instance, we observe that for visibility of $0.994$ the number of one-sided random bits that can be certified is already less than unity. The obtained theoretical results are nevertheless encouraging for the near future, as sources with better visibilities should allow to certify higher and higher amounts of genuine random bits.
\
\\
\

\section{Conclusion}
\label{sec.conclusion}
In conclusion, we show that a recently introduced method for certification of entanglement when semi-device independent measurements realized in a quantum network can be successfully employed {to certify all types of entanglement from a three-qubit W state. Furthermore, such semi-DI entanglement certification is achieved in all tripartite steering scenarios, and without the need to consider different measurements among different scenarios}. We study in detail the case of a tripartite configuration, even though the method is valid for larger networks. Entanglement is witnessed for a few illustrative combinations of trusted and untrusted measurements. We also present experimental results obtained with a
proof of principle optical  set-up. We observe good qualitative agreement between theory and experimental results, and verify the strong dependence of the witnessed entanglement on the degree of purity of the initial state. 

We further investigate the certification of genuine randomness. We observe that one-sided randomness cannot be certified. However, considering bipartitions of
the tripartite W state and the bipartition with two elements untrusted, it is possible in principle to certify up to 1.58 random bits. Analyzing the experimental implementation, 
 $0.26 \pm 0.04$ random bits were certified. {This in fact constitutes the first experimental demonstration of one-sided device-independent randomness certification}. This discrepancy between the expected and measured values emphasizes the critical dependence of the amount of random information certified on the purity of the initial state.  Our results promote the W state as a key candidate for the implementation of device semi independent protocols, where some of the parties use untrusted devices.

\section{Acknowledgements}
We thank O. Jimenez-Farias for discussions. This work was supported by the mexican CONACyT graduate fellowship program, the Ramon y Cajal fellowship, Spanish MINECO (Severo Ochoa grant SEV-2015-0522 and FoQus), the AXA Chair in Quantum Information Science, the Generalitat de Catalunya (SGR875), Fundacio Privada Cellex, ERC CoG QITBOX and the ERC AdG NLST. Financial support was also provided by Brazilian agencies FAPERJ, CNPq, CAPES, and the National Institute of Science and Technology for Quantum Information.

\begin{appendix}

\section{Experimental state}
\label{app.exp}
\begin{figure}
\centering
\includegraphics[width = 0.8\linewidth]{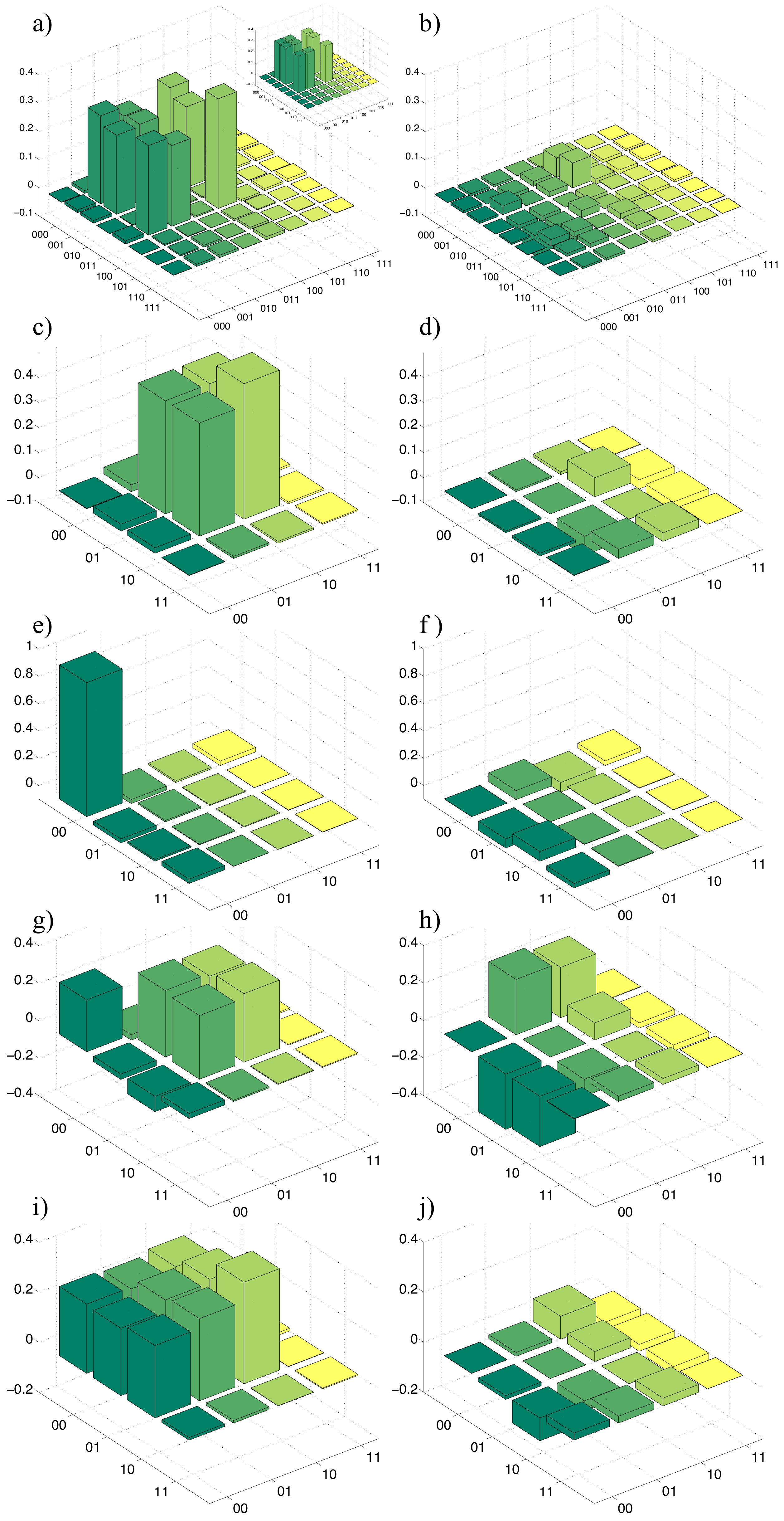}
\caption{{ Full density matrix and examples of reconstructed
conditional density matrices for $\rho_{AC}$ when
 party B uses untrusted measurements. 
The real and imaginary parts of the density matrices are
shown in the left and right columns, respectively. 
Plots a) and b) show the full density matrix $\rho_E$,  
c) and d) show $\rho_{AC}$ when B projects onto $|0\rangle$,
e) and f) show $\rho_{AC}$ when B projects onto $|1\rangle$, 
g) and h) show $\rho_{AC}$ when B projects onto $|R\rangle = (|0\rangle + i |1\rangle)/\sqrt{2}$,
i) and j) show $\rho_{AC}$ when B projects onto $|R\rangle = (|0\rangle + |1\rangle)/\sqrt{2}$. }}
\label{fig:expresults}
\end{figure}

Using quantum state tomography, we reconstruct the experimental density matrix $\rho_E$, shown in Fig.  \ref{fig:expresults}a), to check the quality of the $W$ state produced. We found that purity $\text{P}=\text{Tr}\left({\rho_E}^2\right)$ of the measured state is $0.88 \pm 0.02$ while its fidelity $\text{F}_W=\bra{W} \rho_E \ket{W}$  with respect to the ideal $W$ state is $0.92 \pm 0.01$. The errors in $\text{P}$ and $\text{F}_W$ were calculated using Monte Carlo simulation assuming Poissonian coincidence counting statistics.

The degree of purity and the fidelity can be considered high for a tripartite system, when compared to other physical realisations of multipartite states. Even comparing with other photonic setups, our experiment  produce better quality three-qubit states with larger count  rates. For instance, in Ref.Ê \cite{Kiesel07} the W state is achieved using four photons produced by SPDC. In that experiment the fidelities attained are up to $0.88$ with rates of coincidences 1000 times smaller than in our experiment.  These characteristics highlight the reliability of using two degrees of freedom of a pair of photons rather than three or four photons, as a platform to test theoretical strategies related to three and four-partite entangled states.

Concerning the purity of the experimental bipartite entangled states, which affects the visibility of the $W$ states, we notice that considerably high purities in the range of $0.99$ have been achieved for bipartite states. For increasing the number of parties, obtaining experimental entangled states with such high purity is increasingly challenging, since the purity decreases exponentially with the number of parties in the presence of noise. The present realisation with purity and fidelity to the $W$ state in the range $0.9$ is enough to demonstrate some randomness certification, and to highlight the sensitivity of this task to the visibility of the $W$ state.

 The discussed experimental imperfections also imply a discrepancy at the level of the steering witnesses, which are available online at: https://github.com/mattarcon2tes/Steering. More precisely, the differences which may be found on the experimental witnesses with respect to the theoretical ones ---obtained from pure assemblages of the W state--- is due to the non-unit purity and fidelity of the experimental state, but also is due to the signaling character of the experimental data, which forces one to to consider a non-signalling assemblage to extract the inequality, as explained in \secref{sec.considerations}.
\
\\
\


\end{appendix}

\end{document}